\documentclass[thmsa]{article}

\input tcilatex

\begin{document}

\author{Babur M. Mirza\thanks{
E-mail: bmmirza2002@yahoo.com} \\
Department of Mathematics, Quaid-i-Azam University, \\
Islamabad. Pakistan. 45320}
\title{Travelling Magnetic Waves due to Plasma Surrounding a Slow Rotating
Compact Gravitational Source}
\date{February 4, 2004}
\maketitle

\begin{abstract}
The magnetic field due to an axially symmetric, hot and highly conducting
plasma, taken as an ideal magnetohydrodynamic fluid, surrounding a slow
rotating compact gravitational object is studied within the context of
Einstein-Maxwell field equations. It is assumed that whereas the plasma is
effected by the background spacetime it does not effect the spacetime
itself. The Einstein-Maxwell equations are then solved for the magnetic
field in a comoving frame with the background spacetime described by the
slow rotating Kerr black hole spacetime. It is found that the solutions are
magnetic waves travelling along the azimuthal angle with velocity equal to
the angular velocity of a free falling intertial frame. These general
solutions, when applied to various particular cases of physical interest,
show that for a fixed value of the azimuthal angle the magnetic field is
completely induced by the dragging of the background spacetime.
\end{abstract}

\section{Introduction}

The behavior of plasma in Schwarzschild or Kerr spacetime has been a subject
of increasing interest in recent years. This is mainly due to the fact that
in various astrophysical situations, such as in the magnetospheres of a
strong gravitational source, mostly present in active galactic nuclei
(AGNs), matter exists in a highly ionized state\cite{1,2}. Under such
circumstances the surrounding plasma possesses a strong magnetic field which
not only plays important role in various emission mechanisms but also
effects the dynamics of charged particles and accretion of the surrounding
plasma\cite{3,4,5}. On the other hand the presence of a strong gravitational
source in such cases moulds the background spacetime to produce effects
unknown to the Newtonian theory of gravitation. Most of the attempts to
treat this phenomenon in Schwarzschild or Kerr spacetime either ignore the
presence of strong magnetic field present in such situations or do not
discuss the problem within the frame-work of Einstein-Maxwell equations as
the basic governing equations for the behavior of plasma and the associated
magnetic field. Except under some very simplifying assumptions, the problem
becomes not only mathematically intractable but also offers difficulties for
any general physical interpretation of the results. However, it is found
that it is possible to discuss the problem for the magnetic field, within
the context of Einstein-Maxwell equations, at least for the case of a highly
conducting, axially symmetric plasma surrounding a slow rotating compact
gravitational source, such as a slow rotating neutron star or a slow
rotating black hole.

In this paper we consider the phenomenon to be consisting of an axially
symmetric plasma surrounding a slow rotating compact gravitational source;
taking the plasma as an ideal magnetohydrodynamic (MHD) fluid. The plasma is
assumed to produce no substantial perturbations to the background spacetime
geometry but is taken to be strongly effected by it. We then construct, in
Section 2, the general form of the Einstein-Maxwell equations for the
electromagnetic field outside the plasma in an axially symmetric spacetime.
The formulation is based on a general definition of the electromagnetic
field tensor for an ideal MHD fluid in a curved four dimensional spacetime .
We then explicitly write the field equations in a comoving frame of
reference in slow rotating Kerr black hole spacetime. In Section 3 we obtain
analytical expressions for the magnetic field in the comoving frame under
the condition of perfect conductivity. The system of coupled quasi-linear
partial differential equations thus obtained is shown to admit travelling
wave solutions. These solutions are obtained and discussed using the method
of characteristics which is particularly useful for solving quasi-linear
partial differential equations in comoving frames. The physical significance
of these solutions is then discussed in Section 4. We particularly focus on
those effects which result as the frame dragging of the background
spacetime. It is found that outside the plasma a magnetic field is induced
which for a given azimuthal is due to the dragging of the free falling
inertial frames. The paper concludes, in Section 5, with a summary of these
observations and suggestions for further investigations.

\section{Formulation of the Basic Equations}

\subsection{Preliminaries}

Since we are to describe the phenomenon in a curved four dimensional
spacetime we first need to give a general definition of the electromagnetic
field tensor for an ideal MHD fluid in such a spacetime. Throughout we
assume the gravitational units in which $G=1=c$.

Let the geometry of the spacetime is given by the four dimensional line
element $ds$ defined by

\begin{equation}
ds^2=g_{\alpha \beta }dx^\alpha dx^\beta =g_{tt}dt^2+2g_{t\varphi
}dtd\varphi +g_{\varphi \varphi }d\varphi ^2+g_{rr}dr^2+g_{\theta \theta
}d\theta ^2,\   \tag{1}
\end{equation}
where the metric tensor components $g_{\alpha \beta }$ are independent of
the time coordinate $t$ and the azimuthal angle $\varphi $. The Greek
indices denote coordinates $t,r,\theta ,$ and $\varphi $. We note that the
electromagnetic field tensor $F_{\alpha \beta }$ is a 2-form, that is, a
skew-symmetric tensor field of order two in a spacetime $V_4$. Now if $\eta
_{\alpha \beta \gamma \delta }$ is the volume 4-form of $V_4$ and $u^\alpha $
is a unit vector at an event $x\epsilon V_4$ defining a time like direction
at this point then the electromagnetic field tensor for the MHD fluid is
give by the covariant expression\cite{6,7} 
\begin{equation}
F_{\alpha \beta }=u_\alpha E_\beta -u_\beta E_\alpha +\eta _{\alpha \beta
\gamma \delta }u^\gamma B^\delta ,  \tag{2}
\end{equation}
and similarly in a contravariant form\qquad 
\begin{equation}
F^{\alpha \beta }=u^\alpha E^\beta -u^\beta E^\alpha +\eta ^{\alpha \beta
\gamma \delta }u_\gamma B_\delta ,  \tag{3}
\end{equation}
where the four vectors $E_\alpha $ and $B_\alpha $, denoting the electric
and magnetic field components in the four dimensional spacetime, are
orthogonal to the velocity four vector $u^\alpha $. Here the volume element
4-form of $V_4$ namely $\eta _{\alpha \beta \gamma \delta }$ and its dual $%
\eta ^{\alpha \beta \gamma \delta }$ is defined as\cite{8} :

\begin{equation}
\eta _{\alpha \beta \gamma \delta }=\sqrt{-g}\epsilon _{\alpha \beta \gamma
\delta },\quad \eta ^{\alpha \beta \gamma \delta }=-\frac 1{\sqrt{-g}%
}\epsilon ^{\alpha \beta \gamma \delta },  \tag{4}
\end{equation}
where $g$ represents the determinant of the metric tensor $g_{\alpha \beta }$
and $\epsilon _{\alpha \beta \gamma \delta }$ is the Levi -Civita symbol,
which is $+1,-1,$ and $0$ for a cyclic, anti-cyclic, and non-cyclic
permutation of $\alpha \beta \gamma \delta $ respectively. It should be
noticed that the choice of spacetime $V_4$ is quite general here, namely of
a four dimensional vector space, or more generally of a differentiable
manifold of four dimensions. Also the signature of an axially symmetric
metric defined on this manifold must be either (+ - - -) or (- + + + ).

It can be easily shown that the assumption of an everywhere finite $J^\alpha 
$ leads to the fact that in a comoving frame $E^\alpha =0=E_\alpha $\cite{9}%
. In our case, that is outside a hot and highly conducting plasma, the
assumption of everywhere finite current four vector is satisfied, so the
electromagnetic field tensor takes the form

\begin{equation}
F_{\alpha \beta }=\sqrt{-g}\epsilon _{\alpha \beta \gamma \delta }u^\gamma
B^\delta ,  \tag{5}
\end{equation}
and in contravariant form

\begin{equation}
F^{\alpha \beta }=-\frac 1{\sqrt{-g}}\epsilon ^{\alpha \beta \gamma \delta
}u_\gamma B_\delta .  \tag{6}
\end{equation}

\subsection{The Einstein-Maxwell Field Equations for Plasma Surrounding a
Rotating Kerr Black Hole}

If we interpret $g_{\alpha \beta }$ as the potentials of a gravitational
field determined by the Einstein field equations then we can write the
electromagnetic field equations for the MHD fluid in a spacetime whose
geometry is given by the general relativistic field equations. The resulting
set of equations, called Einstein-Maxwell field equations, can be expressed
as $\qquad $%
\begin{equation}
F_{\alpha \beta ,\gamma }+F_{\beta \gamma ,\alpha }+F_{\gamma \alpha ,\beta
}=0,  \tag{7}
\end{equation}

\begin{equation}
\left( \sqrt{-g}F^{\alpha \beta }\right) _{,\beta }=4\pi \sqrt{-g}J^\alpha ,
\tag{8}
\end{equation}
where the electromagnetic field tensor, in covariant and contravariant form,
is given by expressions (5) and (6). Also here, as usual, $,\alpha $ as a
subscript denotes partial derivative with respect to the coordinate $%
x^\alpha $.To discuss these equations in full we must choose the four
velocity vector which means that we must specify the comoving observer. A
convenient and physically meaningful choice of the comoving observer is the
one belonging to the class of observers having zero angular momentum (ZAMO)%
\cite{10}. This is an observer, circling the Kerr black hole with angular
velocity $\omega $ at a fixed $r$ and $\theta $, being brought into motion
by the dragging of the background spacetime. Thus for a ZAMO $u_r$ and $%
u_{\theta \text{ }}$ vanish and the velocity four vector is given by $%
(u^t,0,0,u^\varphi )$. Moreover, since the ZAMO defines a comoving frame
therefore, the electric field vanishes also. Under these conditions we
obtain the following set of Einstein-Maxwell field equations for the first
pair 
\begin{equation}
(\sqrt{-g}u^tB^r)_{,r}+(\sqrt{-g}u^tB^\theta )_{,\theta }+(\sqrt{-g}%
u^tB^\varphi )_{,\varphi }=0,  \tag{9}
\end{equation}
\begin{equation}
(\sqrt{-g}u^\varphi B^\theta )_{,\theta }+(\sqrt{-g}u^tB^\varphi )_{,t}+(%
\sqrt{-g}u^\varphi B^r)_{,r}=0,  \tag{10}
\end{equation}
\begin{equation}
(\sqrt{-g}u^\varphi B^\theta )_{,\varphi }+(\sqrt{-g}u^tB^\theta )_{,t}=0, 
\tag{11}
\end{equation}
\begin{equation}
(\sqrt{-g}u^\varphi B^r)_{,\varphi }+(\sqrt{-g}u^tB^r)_{,t}=0,  \tag{12}
\end{equation}
and similarly from expression (8) the second pair is

\begin{equation}
(u_\varphi B_\theta )_{,r}+(u_\varphi B_r)_{,\theta }=4\pi \sqrt{-g}J^t, 
\tag{13}
\end{equation}
\begin{equation}
-(u_\varphi B_\theta )_{,t}+(u_tB_\varphi )_{,\theta }-(u_tB_\theta
)_{,\varphi }=4\pi \sqrt{-g}J^r,  \tag{14}
\end{equation}
\begin{equation}
-(u_\varphi B_r)_{,t}-(u_tB_\varphi )_{,r}+(u_tB_r)_{,\varphi }=4\pi \sqrt{-g%
}J^\theta ,  \tag{15}
\end{equation}
\begin{equation}
(u_tB_\theta )_{,r}-(u_tB_r)_{,\theta }=4\pi \sqrt{-g}J^\varphi .  \tag{16}
\end{equation}

\section{Travelling Wave Solutions to the Einstein-Maxwell Field Equations}

Given the current four vector $J^\alpha $, the metric tensor $g_{\alpha
\beta }$, and the velocity four vector $u^\alpha $ the set of equations
(9)-(16) can solved numerically, to give the magnetic field as a function of 
$(t,r,\theta ,\varphi )$ provided that the problem is well posed and the
solution exists. However it is very difficult to solve the system of these
quasi-linear partial differential equations analytically in most cases
especially for a general definition of $J^\alpha $.

Now since plasma, as it usually exists in compact star magnetospheres, is
highly conducting; it follows from the generalized Ohm's law that

\begin{equation}
J^\alpha =\sigma u^\alpha ,  \tag{17}
\end{equation}
where $\sigma $ is the charge density in the ZAMO. Since outside the plasma $%
\sigma $ is zero, therefore $J^\alpha $ vanishes outside the plasma
surrounding the Kerr black hole The case when $J^\alpha =0$ is not only
physically meaningful but can also be used as a first approximation to more
general cases, such as the magnetic field inside the magnetosphere of a
compact gravitational source. Further we assume that the spacetime is the
slow rotating Kerr metric given by

\begin{equation}
ds^2=-e^{2\Phi (r)}dt^2+e^{-2\Phi (r)}dr^2+r^2d\theta ^2+r^2\sin ^2\theta
d\varphi ^2-2\omega (r)r^2\sin ^2\theta d\varphi dt,  \tag{18}
\end{equation}
where

\begin{equation}
e^{2\Phi (r)}\equiv (1-\frac{2M}r),  \tag{19}
\end{equation}
$M$ is the mass of the compact star and 
\begin{equation}
\omega (r)\equiv \frac{d\varphi }{dt}=\frac{2J}{r^3},  \tag{20}
\end{equation}
is the angular velocity of a free falling intertial frame, whereas $J$ is
the total angular momentum of the Kerr black hole as measured from infinity%
\cite{11}. Finally, under the assumption of slow rotation (i.e., $\omega
(r)^2$ and higher powers are neglected), we take explicitly the velocity
four vector in ZAMO as follows:

\begin{equation}
u^\alpha =(u^t,0,0,\omega u^t),\quad u_\alpha =(-\frac 1{u^t},0,0,0), 
\tag{21}
\end{equation}
where 
\begin{equation}
u^t=e^{-\Phi (r)}.  \tag{22}
\end{equation}
Under these assumptions we find that the pair of equations (13)-(16) is just
the condition of integrability i.e., the partial derivatives satisfy the
condition $\partial _\alpha \partial _\beta =\partial _\beta \partial
_\alpha $ for $\alpha \neq \beta $. Here we assume, as usual, that at least
in the region outside the plasma all relevant physical quantities are
differentiable and single valued. It is worth pointing out here that when
searching for shock wave solutions to Einstein-Maxwell equations this
assumption may not be true\cite{12}. When the integrability condition is
satisfied the non-identical set of Einstein-Maxwell equations reduces to
equations(9)-(12), which can be further simplified to give

\begin{equation}
\frac 1{u^t}(\sqrt{-g}u^tB^r)_{,r}+(\sqrt{-g}B^\theta )_{,\theta }+\sqrt{-g}%
B^\varphi {}_{,\varphi }=0,  \tag{23}
\end{equation}
\begin{equation}
\sqrt{-g}B^\varphi {}_{,t}+\frac 1{u^t}(\sqrt{-g}\omega u^tB^r)_{,r}+\omega (%
\sqrt{-g}B^\theta )_{,\theta }=0,  \tag{24}
\end{equation}
\begin{equation}
B_{,t}^\theta +\omega B^\theta {}_{,\varphi }=0,  \tag{25}
\end{equation}
\begin{equation}
B^r{}_{,t}+\omega B_{,\varphi }^r=0,  \tag{26}
\end{equation}
where $\omega $ is given by (20) and $u^t$ is given by (22).

Differentiating with respect to $r$ the second term in equation (24) as a
product of $\omega (r)$ and $\sqrt{-g}u^tB^r$, and using equation (23) in
the resulting expression we obtain a single partial differential equation

\begin{equation}
B^\varphi {}_{,t}+\omega B^\varphi {}_{,\varphi }=-B^r\omega _{,r}.  \tag{27}
\end{equation}

Now to solve equation (25), (26), and (27) we apply the method of
characteristics. Suppose $f(t,\varphi )$ represents any of the component $%
(B^r,B^\theta ,B^\varphi )$, then we have

\begin{equation}
\frac{df}{dt}=f_{,t}+f_{,\varphi }\frac{d\varphi }{dt}.  \tag{28}
\end{equation}

Comparing , say equation (27) and (28), and setting $f=B^\varphi $ we obtain
two simultaneous ordinary differential equations

\begin{equation}
\frac{dB^\varphi }{dt}=-B^r\omega _{,r},  \tag{29}
\end{equation}
and notably

\begin{equation}
\frac{d\varphi }{dt}=\omega .  \tag{30}
\end{equation}
This relation implies that the characteristics travel at a speed $\omega $.
Solving (30) we get the characteristics $\varphi =\omega t+\varphi _0$. On
the other hand equation (31) has solution $B^\varphi =-B^r(\varphi
_0)t\omega _{,r}+h(\varphi _0)$. Combining the two results we have the
general solution to equation (27):

\begin{equation}
B^\varphi {}=-B^r(\varphi -\omega t)t\omega _{,r}+h(\varphi -\omega t), 
\tag{31}
\end{equation}
where the function $h(t,\varphi )$ satisfies the condition

\begin{equation}
h_{,t}+\omega h_{,\varphi }=0.  \tag{32}
\end{equation}

A similar procedure for equations (25) and (26) shows that $B^r$ and $%
B^\theta $ indeed have general solutions of the form $f(\varphi -\omega t)$,
where $f$ is an arbitrary function of its argument $\varphi -\omega t$, and
whose form is determined by the physical constraints.

\section{Physical Interpretation of the Travelling Magnetic Wave Solutions}

In the preceding Section it was shown that the magnetic field outside a
plasma surrounding a slow rotating black hole has a general travelling wave
solution of the type

\begin{equation}
B^r=f(\varphi -\omega t),\quad B^\theta =g(\varphi -\omega t),\quad
B^\varphi =-ft\omega _{,r}+h(\varphi -\omega t)  \tag{33}
\end{equation}
where $f$, $g$, and $h$ are arbitrary functions of the argument $\varphi
-\omega t$ and where $\omega $ given by (30) can be interpreted as the
velocity with which a characteristic moves or the velocity of a wave profile
. To to further clarify the physical significance of these solutions, let us
discuss some specific cases of $f$, $g$, and $h$ and their dependence on $%
\omega $.

\subsection{Plane Wave Solution}

If the plasma perturbations are such that the magnetic field remains
stationary outside the plasma, then it follows from (33) that the magnetic
field components are given by

\begin{equation}
B^r=\varphi -\omega t,\quad B^\theta =\varphi -\omega t,\quad B^\varphi
=-(\varphi -\omega t)t\omega _{,r},  \tag{34}
\end{equation}
where for simplicity we have taken $h=0$. Furthermore $\varphi $ being the
azimuthal angle can be given a fixed value, say zero. We notice that for a
fixed value of $\varphi $, the magnetic field components depend directly on
the frame dragging frequency $\omega $. This means that the magnetic field,
in this case, is induced by the rotation of the Kerr black hole. However
this induced field may not last for long as the assumption of being
stationary is not valid for actual compact star magnetospheres where plasma
is usually in a state of high oscillations.

\subsection{Oscillatory Solution}

For an oscillating plasma the functions $f$, $g$, and $h$ have a sinusoidal
form represented by $\sin $ or $\cos $ functions. As before we take $h=0$,
and $\varphi =0.$ Then for the magnetic field components we have from (33):

\begin{equation}
B^r=\cos \omega t,\quad B^\theta =\sin \omega t,\quad B^\varphi =-t\omega
_{,r}\cos \omega t.  \tag{35}
\end{equation}

Here we notice the frequency of oscillations is the dragging frequency. The
magnetic field in this case is again induced by rotating gravitational
source. We notice that the field components $B^r$ and $B^\theta $ remain
bounded for all time, however the component $B^\varphi $ is unbounded as $t$
increases. For actual physical situations the magnetic field must decay with
the passage of time. Again the oscillatory solutions are not valid for long
periods time for a real physical situation.

\subsection{Exponentially Decaying Solution}

The solutions (33) to the Einstein-Maxwell equation, for $h=0$ and $\varphi
=0$, can be written in the form of exponentially decaying functions as

\begin{equation}
B^r=\exp (-\omega t),\quad B^\theta =\exp (-\omega t),\quad B^\varphi
=-t\omega _{,r}\exp (-\omega t);  \tag{36}
\end{equation}
or in terms of complex representation

\begin{equation}
B^r=\exp (-i\omega t),\quad B^\theta =\exp (-i\omega t),\quad B^\varphi
=-t\omega _{,r}\exp (-i\omega t).  \tag{37}
\end{equation}

Both of these solutions are bounded for very large time scales and each can
be used to represent the magnetic field in the case of a highly conducting
plasma surrounding a slow rotating Kerr black hole.

The above examples show that the magnetic field, for a given value of $%
\varphi $, is induced by the rotating gravitational source and it decreases
with distance from the source. Furthermore the magnetic field is bounded
when it has an exponentially decaying form. We can interpret this as follows:%
\textit{\ For a compact gravitational source possessing a highly ionized
atmosphere there are magnetic waves travelling along the azimuthal with
velocity equal to the angular velocity of a free falling inertial frame
induced by the rotating gravitational source and which remains bounded as a
decaying exponential function of time.}

\section{Discussion}

In this paper we have investigated the solutions for the magnetic field
components admitted by the Einstein-Maxwell field equations in a comoving
frame (ZAMO). Taking the case of plasma as an ideal MHD fluid surrounding a
slow rotating Kerr black hole, we found that the magnetic field can be
written as a travelling wave moving with velocity $\omega $ in the azimuthal
direction. By considering various cases of physical importance we observed
that for a fixed $\varphi $ the magnetic field depends directly on the frame
dragging frequency $\omega $, thus we can regard the magnetic field to be
induced, at least partially, by the spacetime dragging. There is an
indication\cite{13} of such induced magnetic field in the gravitomagnetic
approximation to the general theory of relativity\cite{14}, however this is
the first indication of such an effect as a consequence of the
Einstein-Maxwell field equations. Since the induced magnetic field depends
on the frame dragging frequency, its magnitude is extremely small to be
directly measurable. In extreme astrophysical situations where such an
effect is plausibly observable the correct strategy will be to study the
effects of the induced magnetic field on various physical processes, such as
the accretion of charge particles in a compact star magnetosphere. These and
other features of the travelling wave solutions to the Einstein-Maxwell
field equations are under considerations, which we hope to report in near
future.

\textbf{Acknowledgments}

I am grateful to Dr. A. Qadir and Dr. B. J. Ahmedov for useful comments and
suggestions.

\end{document}